# Sensitivity to initial conditions in an extended activator-inhibitor model for the formation of patterns

R. Piasecki, W. Olchawa, K. Smaga

Institute of Physics, University of Opole
Oleska 48, 45-052 Opole, Poland

Despite simplicity, the synchronous cellular automaton [D.A. Young, Math. Biosci. **72**, 51 (1984)] enables reconstructing basic features of patterns of skin. Our extended model allows studying the formatting of patterns and their temporal evolution also on the favourable and hostile environments. As a result, the impact of different types of an environment is accounted for the dynamics of patterns formation. The process is based on two diffusible morphogens, the short-range activator and the long-range inhibitor, produced by differentiated cells (DCs) represented as black pixels. For a neutral environment, the extended model reduces to the original one. However, even the reduced model is statistically sensitive to a type of the initial distribution of DCs. To compare the impact of the uniform random distribution of DCs (R-system) and the non-uniform distribution in the form of random Gaussian-clusters (G-system), we chose inhibitor as the control parameter. To our surprise, in the neutral environment, for the chosen inhibitor-value that ensures stable final patterns, the average size of final G-populations is lower than in the R-case. In turn, when we consider the favourable environment, the relatively bigger shift toward higher final concentrations of DCs appears in the G. Thus, in the suitably favourable environment, this order can be reversed. Furthermore, the different critical values of the control parameter for the R and the G suggest some dissimilarities in temporal evolution of both systems. In particular, within the proper ranges of the critical values, their oscillatory behaviours are different. The respective temporal evolutions are illustrated by a few examples.

## 1. Introduction

A large variety of spatial patterning can be observed in nature. Full understanding the dynamics of spatio-temporal patterns is still an interesting theoretical problem. For the pattern formation, which is temporally stationary, reaction-diffusion processes are basic mechanisms in the famous Turing model [1]. He showed that under certain conditions, a pair of reacting and diffusing chemicals called morphogens could produce steady state heterogeneous spatial patterns of chemical concentration. Since Turing's seminal paper, numerous non-linear models based on his original idea have been explored. For example, the book by Meinhardt [2] is devoted to applications of the reaction-diffusion model. The fact that the reaction-diffusion model is just a disguised implementation of local autocatalysis with lateral inhibition was first noticed by Gierer and Meinhardt [3]. An elementary mathematical introduction to this field is given in the textbook by Edelstein-Keshet [4]. It gives a broad collection of models for development and pattern formation in spatially distributed biological systems. At more advanced level, the well-known Murray's book [5] provides comprehensive coverage of the diverse mechanisms involved in biological pattern formation. It is worth mentioning also the Bar-Yam's book [6] describing a dynamics of complex systems, and the second one by Ilachinski [7] dealing with a discrete universe from the cellular automata viewpoint. These books provide a valuable introduction into the domain of various methods of patterns formation.

Many models of pattern formation employ the general phenomenon of local instabilities coupled with lateral inhibition. We point out just two of the related brief reviews. The qualitative similarities amongst the models based on local activation with lateral inhibition like neural, diffusion-reaction, mechanical and chemotactic ones are discussed by Oster [8]. The last topic involving cell-chemotaxis (the same cells that secrete a chemoattractant are free to move in response to the chemical gradients they set up) was reviewed by Maini [9]. One more point is worth to mention here. The applicability of Turing approach is not limited to the surface of zero curvature. The problem of pattern formation for Turing systems on a spherical surface has also been addressed, *e.g.* in Refs. [10, 11].

Among other models for the formation of patterns, the cellular automata (CA) approach is particularly suitable for computer simulations. Using simple rules, such models allow creating complex spatial patterns indeed. These kind CA models are catching the attention of physicists because of a possible complex dynamics of temporal evolution, not for biological details of realistic patterns formation. To this group belongs spatially discrete model of growing of vertebrate skin patterns proposed by Young [12]. Although diffusion is not explicitly represented, the mechanism for formation of patterns is that of lateral inhibition: local activation and long-range inhibition [4]; *cf.* Fig. 1 in the next section. Despite its simple logical structure, the model can reproduce basic features of vertebrate skin patterns: spots, stripes or mixed forms. When reduced to a morphogenetic field, the model concept described in the next section provides an algorithm involving on–off deterministic switching of cell differentiation on a substrate that is called here a neutral environment.

The basic question that we consider here is to reveal what dynamic changes in the evolution of this model may occur as a result of environmental alterations measured by a single parameter. A particular focus is given to the question: is the final number of differentiated cells (DCs) sensitive to the type of their initial random spatial distribution? This allows obtaining complementary information in connection with Young's suggestion [12]: "I find that five iterations suffice for convergence to a stable pattern, and that the general form of the final pattern is not sensitive to the initial DC distribution." Our findings indicate that the average size of final DC population is clearly sensitive to the type of an initial configuration of DCs. In addition, the characteristic standard deviations of the distributions of final DC-population sizes for the different types of the environmental conditions can be observed. Moreover, we needed a higher number of iterations to terminate the evolution of subsequent patterns and to obtain a stable final configuration. Interestingly, adopting the Ising model terminology of spin variables in the context of pattern formation, the Young's model can be interpreted as describing magnetic system with interactions that are locally ferromagnetic and long-range antiferromagnetic [6]. Thus, as a model of broad applicability in statistical physics, the Young's cellular automaton with further potential modifications opens up many possibilities for the applied research at relatively low cost.

## 2. Young's model and its extension

The model was developed not for an exact description of reality [12], but rather, by doing some approximations, it provides a simplified description of the complex pattern formation process. According to specific rules described below, an initially uniform random distribution (R) of a given number $n_{init}$ (DCs) of differentiated cells (the DCs are represented as black pixels) in a matrix of undifferentiated cells (the UCs as white pixels) can evolve into a white-black skin pattern. The initial arrangement of DCs on the early embryonic skin is considered as a result of possible slow random process of differentiation in the UC cell population. One can envisage that if the process is specifically biased, then also non-uniform distribution of random Gaussian-clusters (G) build of black pixels can be taken into account as an initial configuration.

Within the Young's approach, only DC cells produce at constant rate two diffusible morphogens of different kinds with a given field values, $w_1$ and $w_2$. The activator $w_1 > 0$ (the inhibitor $w_2 < 0$) has the shorter (longer) range and stimulates the differentiation process (the dedifferentiation one). In turn, the UC cells are passive in this model since they produce no active substances. Using the so-called morphogenetic summary field, Young simplifies the activator-inhibitor diffusion theory proposed originally by Swindale [13].

To perform cellular automaton simulations, we employ a typical square grid $L \times L$ with periodic boundary conditions in both directions. The sum of morphogens, which influences every cell at discrete $(x, y)$ position from all neighbouring DCs decides what fate is of the cell. The original mechanism of patterns formation includes short-range activation $w_1$ for $r_i \leq R_1$ (in the I region) and long-range inhibition $w_2$ for $R_1 < r_i \leq R_2$ (in the II region); *cf*. Fig. 1. The $r_i$ means the radial distance of the $i$th DC from the $(x, y)$-cell. For the model parameters $w_1$, $w_2$, $R_1$ and $R_2$, the rule of time-evolution of every cell, see (2), depends on the summary field $W(x, y; t)$ calculated at time $t$ as follows

$$W(x, y; t) = \sum_{r_i \in I} w_1 + \sum_{r_i \in II} w_2, \qquad (1)$$

where *i* relates to all neighbouring DCs at positions $r_i$ in the regions I and II.

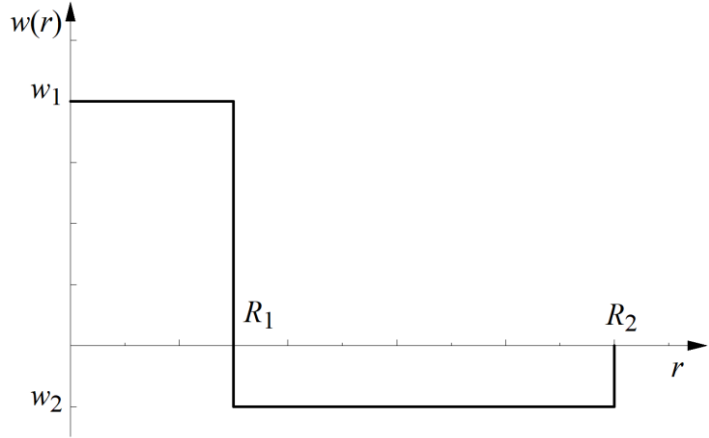

Fig. 1. A discrete activation-inhibition field following the Young's model [12].

Before go further, we recall the conceptually simple extension of the above model. The function $W(x, y; t)$ is directly linked to the effective concentration of the two morphogens at that point and moment *t*. However, the on−off switching of cell differentiations can be also affected by already present chemical or physical properties of the substrate. The substrate material can be equally called "environment". So far, the basic model parameters, $w_1$, $w_2$, $R_1$ and $R_2$ relate to a morphogenetic field given by (1), which is approximated by two linear regions I and II as described in Ref. [12]. The last $\varepsilon$-parameter has been already introduced, although in a different context [14]. It extends the capability of the model making it sensitive to the three general types of the environmental conditions: the favourable ($\varepsilon < 0$), the neutral ($\varepsilon = 0$) and the hostile ($\varepsilon > 0$). Now, for each (*x*, *y*; *t*)-cell the following situations are possible at time $t + 1$:

(a) If $W(x, y; t) < \varepsilon$ then DC(UC) becomes (remains) a UC at time $t + 1$
(b) If $W(x, y; t) = \varepsilon$ then the cell does not change state at time $t + 1$ (2)
(c) If $W(x, y; t) > \varepsilon$ then UC(DC) becomes (remains) a DC at time $t + 1$

If $\varepsilon > 0$ then the actual $W(x, y; t)$ must be a little stronger to change UC into a DC in comparison to the original model [12]. It makes more difficult such changes supporting the lowering of the size of final DC-population. The opposite situation appears for $\varepsilon < 0$. In the case of a neutral environment with $\varepsilon = 0$, its effective influence is negligible by definition, and Young's model is recovered.

Once the results of changing states for each grid cell are saved as a separate subsequent pattern, we consider this moment as the first iteration step $j = 1$. It can be equally named as the step $t = 1$ of temporal evolution.

Thus, the total length of evolution can be measured in iteration steps. Then, the resulting black−white pattern with a current DC-population of size $n(j)$ becomes the new starting configuration. So, within this approach, the update of cells is of synchronous type because, effectively, all the cells can be treated as those updated simultaneously. Denoting the number of "positive" UC → DC and "negative" DC → UC changes in the $j$th iteration by $n^+(j)$ and $n^-(j)$, the iteration process is repeated until $n^+(j) = n^-(j) = 0$. This means that an evolving system reaches a stable configuration that is a final pattern and no longer changes. The related final population size $n_f(DC)$ can be reached either monotonically or, by damped oscillations of a current number of DCs.

However, a kind of unexpected behaviour in temporal evolution can occur with never-ending oscillations of pattern's population sizes. For example, the sustained oscillations between populations of *different sizes* as well as the locally degenerated configurations (local spatial "frustration") with on−off switching black ↔ white but with a *conserved* total number of DCs. The latter very rare cases are not characteristic for the ranges of the model parameters considered in this work and they were omitted. On the other hand, making use of an asynchronous updating of a system, what increases essentially the computation cost, probably such oscillatory behaviour could be eliminated [6]. This point deserves further studies.

### 3. Illustrative examples

As the basic control parameter we choose the $w_2$, which measures the strength of net inhibition effect in II region, while the $\varepsilon$ will be used as the auxiliary parameter that describes the environmental features needed in the modelling. By the $w_2^*(R)$ and the $w_2^*(G)$ we denote the respective "critical" values of the control parameter. For a single run with a given random seed, they indicate the beginning of the so-called oscillatory behaviour of the population size $n(w_2; R)$ or $n(w_2; G)$ calculated as a function of the control parameter. In turn, when the averaged oscillatory behaviour is analysed in each of the systems for 100-run trials, the exact critical values cannot be obtained. Instead, on the corresponding figures only the related approximated values are presented.

The other model parameters are kept fixed in this work, namely a square grid of linear size $L = 83$ (in pixels), $R_1 = 1.5$, $R_2 = 6$, $w_1 = 1$ and the initial number $n_{init} = 455$ of DCs. For the G-systems, a non-uniform initial distribution in form of 65 random Gaussian-clusters with the centres randomly drawn and composed of 7 DCs, the black pixels in each of the clusters are distributed with a standard deviation $\sigma_x = \sigma_y = 1.5$.

When we illustrate dependent on an environment histograms of the final population sizes, the fixed value of $w_2 = -0.08$ is used. Otherwise, the $w_2$ works as the control parameter.

#### 3.1 Creating test patterns

For control purposes, we present first the simplest test-patterns evolving from a single DC cell centrally positioned ($x = 42$, $y = 42$) on a square grid.

The following snapshots taken after 25, 45 and the final step are depicted in Figs. 2 (a) and (b) with the $\varepsilon = 0$ and $\varepsilon = 0.04$, respectively. Both characteristic final patterns show a high symmetry. They can be used to verify the correctness of a CA algorithm.

As expected, in a slightly hostile environment, the final population size $n_f$ is lower than that for the neutral case, which is a typical behaviour. Obviously, the differences in the corresponding patterns become more distinct at the later stages of temporal evolution.

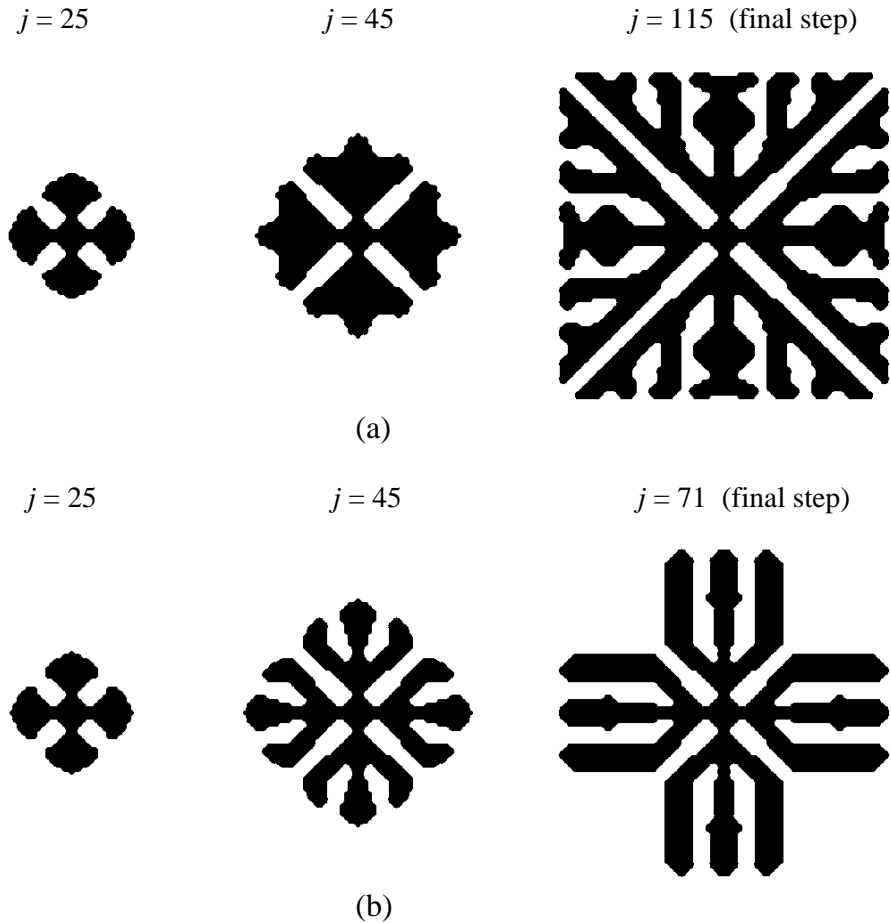

Fig. 2. Test-patterns evolving from the simplest initial configuration consisting of a single DC cell (black pixel) centrally positioned on a square grid of linear size $L = 83$ for $R_1 = 1.5$, $R_2 = 6$, $w_1 = 1$. (a) The neutral environment with $\varepsilon = 0$; (b) The slightly hostile one with $\varepsilon = 0.04$.

### 3.2 Simple examples of stable final patterns for the R- and G-systems

Let us now consider the changes of a current population size $n(j)$ with the fixed value $|w_2| = 0.08 < |w_2^*|$, which ensures a stable final configuration. The following values of environmental parameter are selected, $\varepsilon = -0.5$, 0 and 1. In Ref. [12], a remark about the general form of final patterns is made. The author is probably right in the point that for the different initial random configurations in a neutral environmental conditions ($\varepsilon = 0$), the parameters responsible for the formation only spots never produce solely stripes and reversely. However, for the systems with a different type of an initial distribution as the R-system in Fig. 3 and G one in Fig. 4, a subtle difference can appear, *e.g.* in Fig. 3 (b) left compared with Fig. 4 (b) left. This is related to the spatial inhomogeneity degree as it will be explained in Subsection 3.4. On the other hand, sometimes also a mixed patterning appears; *cf.* Fig. 3 (b) right with Fig. 4 (b) right.

Now we shall illustrate how various environmental conditions influence the formation of pattern for a given type of a system. We expect that the associated various non-zero values of the parameter $\varepsilon$ may change some structural features of the final pattern.

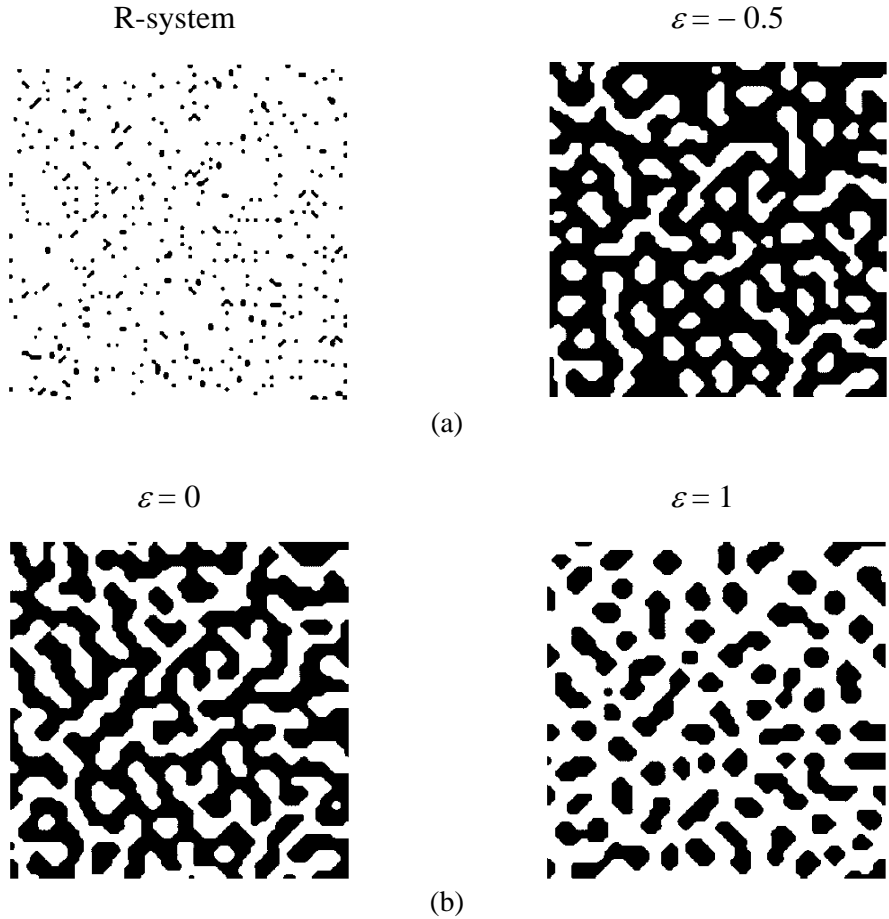

Fig. 3. Patterns produced for $w_2 = -0.08$ but with using the uniform random R-configuration of $n_{init} = 455$ DCs (volume concentration $\varphi_{init} \approx 0.066$). The initial DC number is the same for the next examples until its change is declared. (a) left: The initial R-system. (a) right: The final pattern for a favourable environment. (b) left: The final pattern for a neutral one. (b) right: The final pattern for a hostile one.

Indeed, the change from a stripe in Fig. 3 (a), right to a mixed spot-stripe pattern in Fig. 3 (b), right can be observed for $\varepsilon = -0.5$ and $\varepsilon = 1$, respectively. Similar behaviour can be observed in Figs. 4 (a), right and (b), right. With appropriately hostile the $\varepsilon$-values, one can observe nearly a complete disappearance of DC-population. On the other hand, for favourable enough environment the final population can be over-crowded which relates to an almost black pattern.

Within the range of parameters corresponding to Figs. 3 and 4, the current numbers $n(j)$ evolve in a standard way as Fig. 5 shows. This kind of temporal evolution is a typical one for the original model. The evolution of both R- (the open circles) and G-system (the filled circles) terminates finally with a population size $n_f(R)$ and $n_f(G)$ that strongly depends on the $\varepsilon$ value. As expected, the lowest $n_f$ corresponds to the most hostile environment, that is to $\varepsilon = 1$ in both cases.

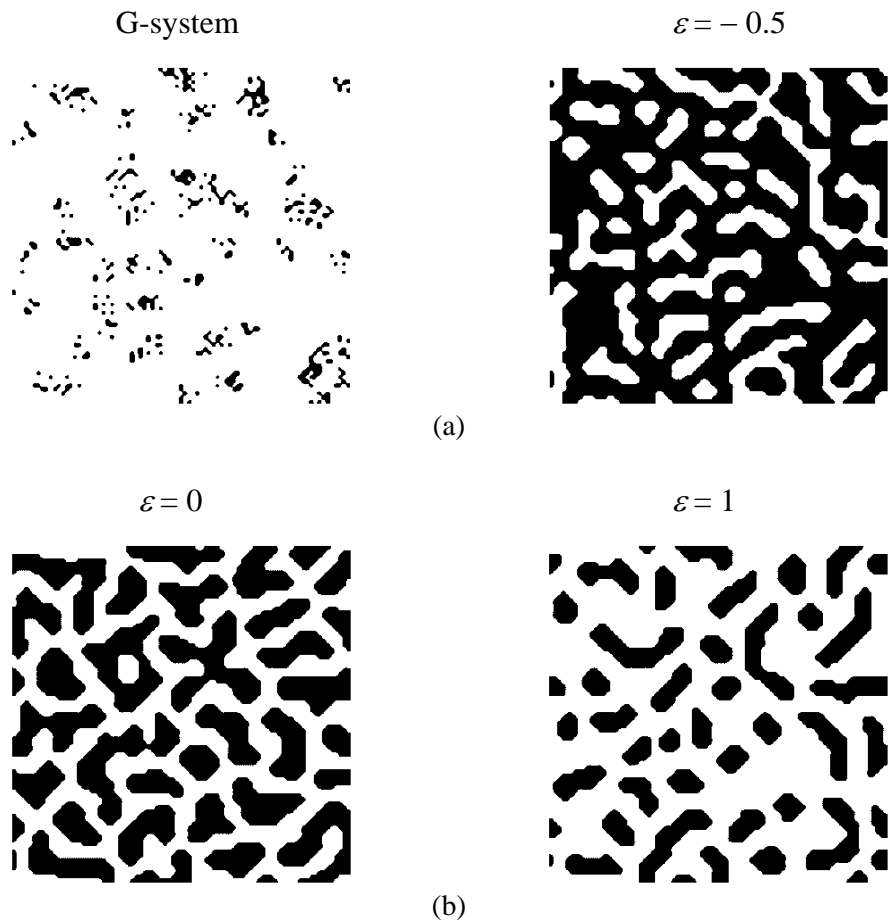

(a)

(b)

Fig. 4. The same as Fig. 3 but for the non-uniform initial distribution in the form of random Gaussian-clusters (the G-system). The initial G-configuration includes 65 clusters with the centres randomly selected. Each of the clusters is composed of 7 DCs. The black pixels in the Gaussian-clusters are distributed with a standard deviation $\sigma_x = \sigma_y = 1.5$.

In the next section, we will exhibit also the statistically significant connection between the average size of a final population and the type of an initial random configuration of DCs what complements the earlier mentioned Young's remark [12].

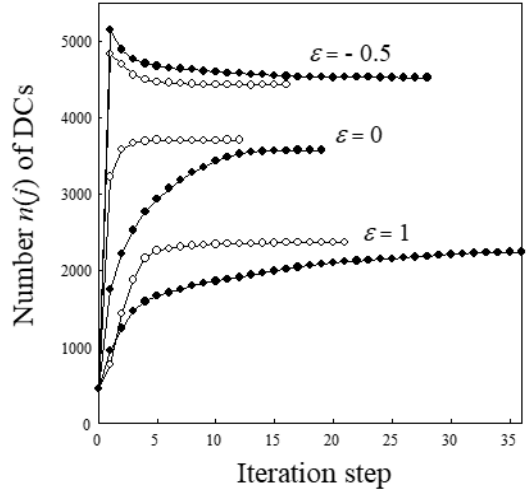

Fig. 5. A current number $n(j)$ of DCs as a function of iteration step for the patterns in Figs. 3, which relate to the initial R-configuration (the open circles). Correspondingly, for the patterns in Figs. 4 that relate to the initial G-configuration (the filled circles). Note the close to monotonic changes of $n(j)$ at the final stages of temporal evolution.

### 3.3 Histograms of sizes of final populations for the G- and R-systems

We have already mentioned that for every type of an initial random distribution of DCs, the size of final population $n_f$ should be statistically sensitive to uncontrollable details of a spatial configuration. Indeed, for 10 000-run trials of G- and R-system the appropriate histograms of $n_f$ can be well fitted by a Gaussian-type function

$$F(n_f) \propto \exp\left[-\frac{(n_f - \tilde{n}_f)^2}{2\sigma^2}\right]. \qquad (3)$$

Moreover, in Fig. 6, we observe that the most probable final population size, denoted here as $\tilde{n}_f$, explicitly depends on a type of the initial distribution. For instance, when $\varepsilon = 0$, the best fit is $\tilde{n}_f(G) = 3610 \div 3611$ with a standard deviation $\sigma(G) \cong 35.3$ and, correspondingly, $\tilde{n}_f(R) = 3682 \div 3683$ with $\sigma(R) \cong 27.6$. In turn, if $\varepsilon = -0.48$, we notice the opposite behaviour. Now, $\tilde{n}_f(G) = 4389 \div 4390$ with $\sigma(G) \cong 62.5$ and correspondingly, $\tilde{n}_f(R) = 4320 \div 4321$ with $\sigma(R) \cong 36.2$. For the middle pair of G- and R-histograms that relate to $\varepsilon = -0.24$, we obtain the best fit for $\tilde{n}_f(G) = \tilde{n}_f(R) = 3917 \div 3918$ with different standard deviations, $\sigma(G) \cong 78.3$ and $\sigma(R) \cong 28.2$.

These observations show that some features of the G-systems leading to a smaller $\tilde{n}_f$, can be over-come in a favourable enough environment. The relatively bigger shifts of the G-histograms (the filled circles) compared with the R-case (the open circles) toward higher final concentrations of DCs, clearly support this conclusion.

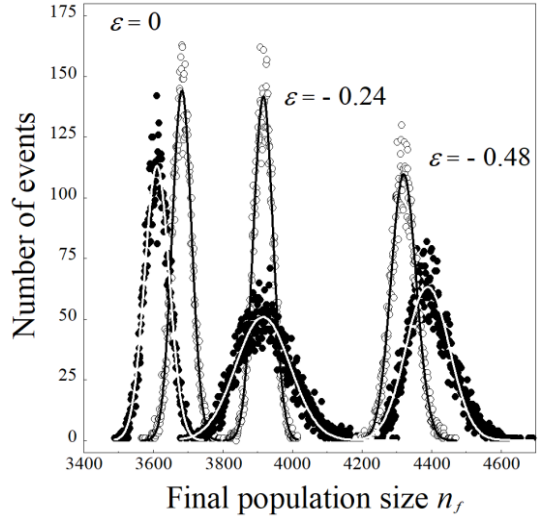

Fig. 6. The histograms of 10 000-run trials for $w_2 = -0.08$ and chosen values of $\varepsilon = 0, -0.24, -0.48$. The filled and the open circles stand for the initial G- and R-configurations, respectively. We depict also the corresponding Gaussian-type fitting functions (the white lines for the G and the black lines for the R, *cf.* (3)). Note the relatively bigger shift of the most probable final population size $n_f(G)$ compared to $\tilde{n}_f(R)$.

### 3.4 A possible correlation between the degrees of spatial disorder detected in the initial and final configurations

In general, the most probable size of final population for a G-system can be smaller, equal or greater than the counterpart for an R-system. For example, in our case the inequality $\tilde{n}_f(G) < \tilde{n}_f(R)$ for $\varepsilon = 0$ is replaced by the reverse one $\tilde{n}_f(G) > \tilde{n}_f(R)$ for $\varepsilon = -0.48$. This suggests that there is a kind of coupling existing between the intensity of environmental alterations and the most probable final population size $\tilde{n}_f$. Moreover, it should be different for each of the types of initial distributions considered in this work. Our previous simulations suggest that this effect is slightly stronger in G-systems.

The type of an environment also influences the length of temporal evolution. The G-systems evolve usually longer in time because of their greater initial spatial disorder in comparison to R-systems. The quantitative evaluation of the spatial inhomogeneity degree can be obtained using a simple entropic measure for finite sized objects (see [15] for binary patterns and [16] for grey-scale ones), its $q$-extensions *à la* Tsallis [17] is given in Refs. [18, 19]. The modified entropic measure can be also widely applied to statistical reconstructions of complex grey-scale patterns [20] and prototypical three-dimensional microstructures [21] with the usage of the decomposable multiphase entropic descriptor [22]. The previous developments and latest applications can be found in [23, 24] and citations therein.

In a few words, the entropic descriptor $S_\Delta = (S_{max} - S)/\chi$ for finite-sized objects (FSOs) quantifies the averaged per cell pattern's spatial inhomogeneity (a measure of configurational non-uniformity) by taking into account the average departure of a system's configurational entropy $S = k_B \ln \Omega$ from its maximum possible value $S_{max} = k_B \ln \Omega_{max}$, where the Boltzmann constant will be set to $k_B = 1$ for convenience. For a given $L \times L$ binary image with $0 < n < L^2$ of the black pixels distributed in square and non-overlapping lattice $\chi$-cells of size $k \times k$, the corresponding formulas can be written as follows [15]

$$\Omega(k) = \prod_{i=1}^{\chi} \binom{k^2}{n_i}, \qquad (4)$$

$$\Omega_{max}(k) = \binom{k^2}{n_0}^{\chi - r_0} \binom{k^2}{n_0 + 1}^{r_0}, \qquad (5)$$

and

$$S_\Delta(k) = \frac{r_0}{\chi} \ln\left(\frac{k^2 - n_0}{n_0 + 1}\right) + \frac{1}{\chi} \sum_{i=1}^{\chi} \ln\left(\frac{n_i!(k^2 - n_i)!}{n_0!(k^2 - n_0)!}\right), \qquad (6)$$

where $\chi = (L/k)^2$, $n_1 + n_2 + \ldots + n_\chi = n$, $n_i \leq k^2$, $r_0 = n \mod \chi$, $r_0 \in 0, 1, \ldots, \chi - 1$ and $n_0 = (n - r_0)/\chi$, $n_0 \in 0, 1, \ldots, k^2 - 1$.

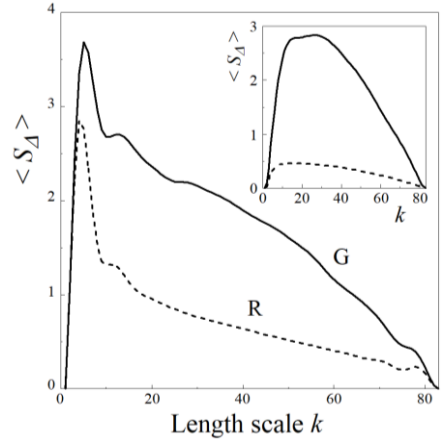

Fig. 7. The average entropic measure $<S_\Delta>$ (*cf.* Eq. (4) in Ref. [15]) *versus* the length scale $k$ in pixels for the sets of initial patterns (see the inset) and the final ones; for the G-system (the solid lines) and for the R-system (the dashed lines). The patterns correspond to the most probable final population sizes with $\varepsilon = -0.48$ and to most frequent length of temporal evolution, $j(G) = 22$ and $j(R) = 11$.

In order to calculate the value of the measure at every length scale $k$, the following property is employed. If the final pattern of size $mL \times mL$, where $m$ is a natural number, is formed by periodical repetition of an initial arrangement of size $L \times L$, then the value of the entropic descriptor at a given length scale $k$ (*commensurate* with the side length $L$) is unchanged under the replacement $L \times L \leftrightarrow mL \times mL$ since it also causes $n \leftrightarrow m^2 n$, $\chi \leftrightarrow m^2 \chi$, $r_0 \leftrightarrow m^2 r_0$ keeping the black phase $\varphi$-concentration, $n_0$ and the corresponding $n_i$ the same.

Now, to overcome the problem of *incommensurate* length scale it is enough to find a whole number *m'* such that *m'L* mod *k* = 0 and replace the initial arrangement of size $L \times L$ by the periodically created one of size $m'L \times m'L$. Then we can define $S_\Delta(k; L \times L, n, \chi) \equiv S_\Delta(k; m'L \times m'L, m'^2 n, m'^2 \chi)$; see the useful properties of the measure indicated in point (6) of [15].

The following evolution rule for every length scale *k* is found: the higher average spatial disorder of an initial population distribution, the higher is an average spatial inhomogeneity of the final pattern, *cf.* Fig. 7. This observation seems to be independent on the values of environmental parameter and true for any pair of the G- and R-systems fulfilling the assumptions about equal initial sizes and comparable final ones. Therefore, we believe that it could be a characteristic feature of the model itself.

### *3.5 The range of parameters encompassing also the oscillatory behaviour*

We would like to present also examples with the oscillatory behaviour during a temporal evolution using the fixed value $\varepsilon = -0.70$ this time. Let us first consider a case of temporal evolution of the G-system with $|w_2| < |w_2^*|$, where $w_2^*$ denotes the critical value of the control parameter. Then damped periodic oscillations of the current population size $n(j; G)$ lead to its well-defined final value. Such a case is shown in Fig. 8 (thick line) for $w_2 = -0.2249$ with $n_f = 1642$ DCs; see the corresponding final pattern in the middle position. However, if $w_2 = -0.2250$, then the temporal evolution shows a totally different dynamics in comparison to the previous one. Now, for $j > 44$, the sustained oscillations of $n(j; G)$ appear. In this case, the two different population sizes are allowable by a system: the upper $n(j; G) = 1826$ DCs while the bottom one equals to 1520 DCs (the filled circles in Fig. 8). It suggests that for given parameters, there is a critical value of the control parameter within the range: $-0.2250 < w_2^*(G) < -0.2249$. The similar behaviour but with the much distinct limit patterns shown at the top and bottom position in Fig. 8 can be found for $w_2 = -0.33$.

We have also investigated the oscillatory behaviour of G-system for other values of $w_2 \in [-1, 0]$ with the step 0.0002. In Fig. 9, we show the values of allowable population sizes $n(w_2)$ as a function of the control parameter. (It should be noted here that for the stable evolution, the population size $n(w_2)$ means the final size $n_f$; otherwise, the $n(w_2)$ denotes the upper or the bottom limit population size, which allow estimating the current range of the related oscillations). In the inset, we clearly observe the beginning of the oscillatory behaviour. The area between the upper and bottom branches has been filled out for a better visualization. The question also arises, is the diagram form of the oscillatory behaviour characteristic one (on average at least) for a given type of initial random configuration of DCs or not?

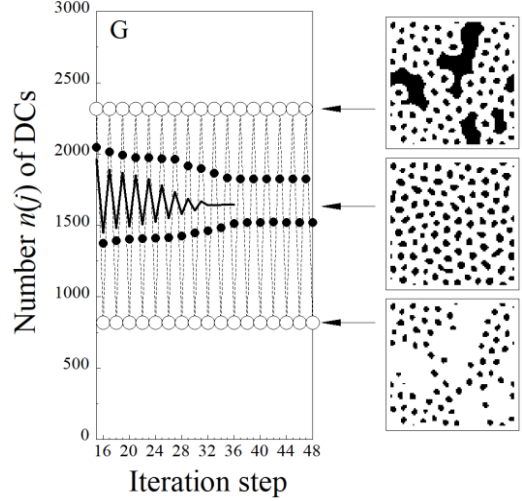

Fig. 8. Evolutionary behaviour of oscillating current population size $n(j; G)$ for the fixed $\varepsilon = -0.7$ and the chosen values of $w_2 = -0.2249$ (the thick line), see the corresponding final pattern (the middle position), $w_2 = -0.2250$ (the filled circles), now the system behaviour is changed to sustained oscillations (for $j > 44$) with a constant amplitude, and $w_2 = -0.33$ (the open circles), see the limit patterns (the top and the bottom positions).

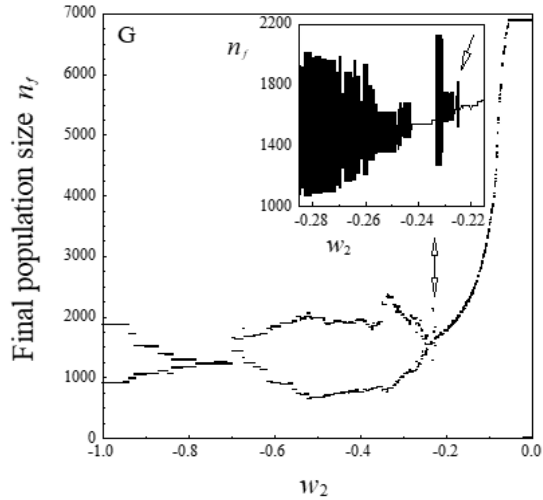

Fig. 9. The corresponding diagram of a single run with the step 0.0002 of $w_2$; in the inset, the enlarged filled area corresponds to the vicinity of the critical value $w_2^*(G)$ indicated by a white arrow.

Before we give below an answer, let us first consider a similar example for R-system. The obtained curves for $w_2 = -0.40$ (thick line), $-0.47$ (filled circles) and $-0.57$ (open circles) are presented in Fig. 10. Now, each of the corresponding temporal evolution process terminates much faster. Also the sustained oscillations of $n(j; R)$ begin earlier than for $n(j; G)$. According to the inset, we are close to the beginning of the oscillatory behaviour in the R-system. We expect that $w_2^*(R) < -0.40$.

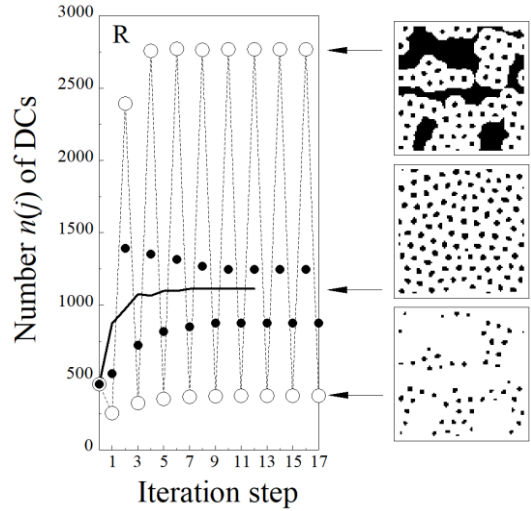

Fig. 10. The same as Fig. 8 but for $n(j; R)$ and different values of $w_2 = -0.40$ (the thick line), $w_2 = -0.47$ (the filled circles), and $w_2 = -0.57$ (the open circles).

Indeed, in Fig. 11, one can observe a diagram of the oscillatory behaviour of different shape from that one for the G-system. The absolute value $|w_2^*(R)| > |w_2^*(G)|$ means a higher sensitivity of the G-system in comparison to the R-system in respect to the oscillatory dynamics.

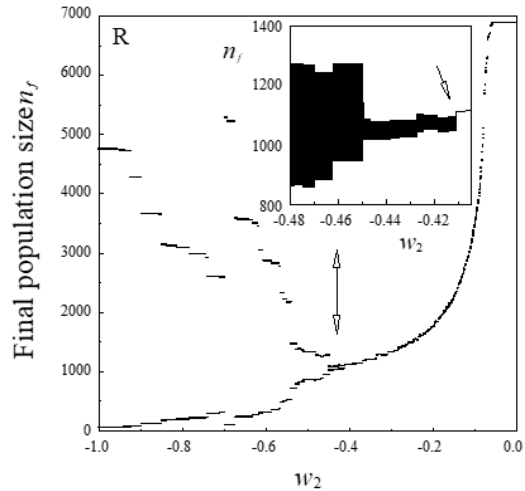

Fig. 11. The same as Fig. 9 but now in the inset we show the vicinity of $w_2^*(R)$. Notice the different shape of the present diagram of the oscillatory behaviour in comparison with the G-system.

The average population sizes $\langle n(w_2; G) \rangle$ and $\langle n(w_2; R) \rangle$ over 100 statistically independent samples as a function of the control parameter $w_2$ with the step 0.01 clearly support this observation, see the solid lines in Fig. 12 and in the inset, respectively. Also, the averaged G-diagram is more compact than the R-diagram but the characteristic shapes of the both diagrams are conserved for the chosen favourable value of the environmental parameter $\varepsilon = -0.7$. Additionally, for a comparison purpose, the case of neutral environment with the $\varepsilon = 0$ is presented (dashed lines).

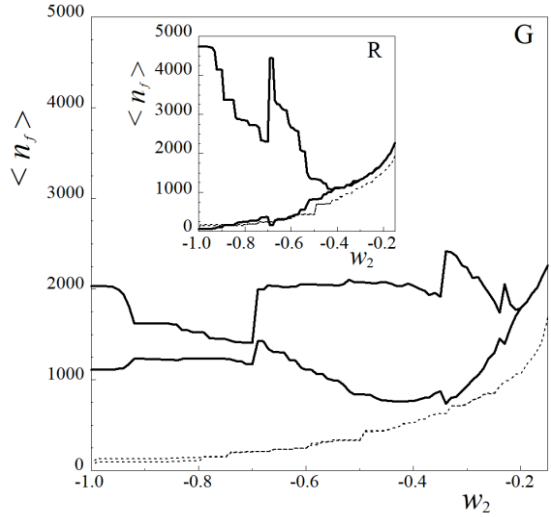

Fig. 12. The averaged oscillatory behaviour in the G-system for 100-run trials with the step 0.01 for $w_2$, a fixed favourable value $\varepsilon = -0.7$ (the solid lines) and for a comparison purpose, a fixed neutral value $\varepsilon = 0$ (the dashed lines). In the inset, the corresponding results are depicted for the R-system. The rescaled similar diagrams, not shown here, are practically independent of a linear size of the system.

To complete the description of the environmental impact, in Figs. 13 (a) and (b), the averages $<n(w_2; G)>$ and $<n(w_2; R)>$ are shown for the selected $\varepsilon$-values. It should be stressed that in the more favourable environmental conditions, the diagrams structure of the oscillatory behaviour in both systems becomes slightly more complex after passing through the related values.

However, the general characteristic features of the diagrams shape in the G and R-systems for different values of the $\varepsilon$-parameter are still preserved. On the other hand, when the G and R-shapes are compared for the same $\varepsilon$-value, the diagrams are essentially different in a form.

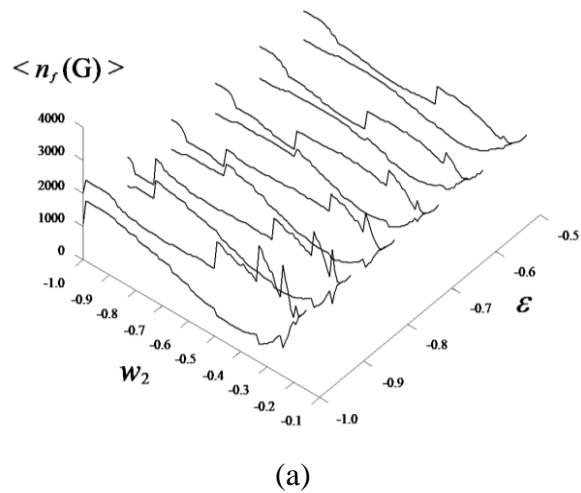

(a)

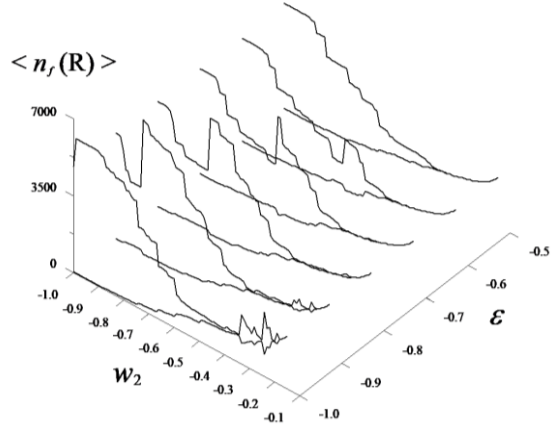

(b)

Fig. 13. Similarly as in Fig. 12, but the averaged oscillatory behaviour is depicted in a series of the diagrams for the chosen favourable values $\varepsilon = -0.5, -0.6, -0.7, -0.8, -0.9$ and $-1$, exclusively. (a) For the G-system. (b) For the R-system.

Finally, a few remarks are in order. Using, for example, the specified set of model parameters: $R_1 = 2$, $R_2 = 3$, $w_1 = 1$, $w_2 = -1$ and $n_{init} = 245$ of DCs on a square grid of linear size $L = 83$, an exotic final pattern containing chessboard parts can be generated out in both the R- and G-systems. The similar type of the symmetrical pattern was a result of the modelling within Monte Carlo approach of the gradual evolution of a variable number of species [25]. Interestingly, according to this model only the better-adapted species show a better ability to organize themselves into symmetrical patterns.

It is worth to notice in this point that in lattice-gas cellular automata such patterns as chessboards are shown to disappear where randomness (a kind of asynchrony) in the updating is added [26]. However, this gives rise to the question, what amount of "asynchrony" is sufficient to destroy such a symmetrical pattern. In the CA model updating context, the authors of [27] emphasize that: "Probably neither a completely synchronous nor a random asynchronous update is realistic for natural systems".

At last, we should also point out for a recently proposed new version of the Turing model [28]. This alternative model is represented by the shape of an activation-inhibition kernel and is named the kernel-based Turing model (KT model). All of it opens a wide field for research topics.

## 4. Concluding remarks

In this work the preliminary results for an extended activator-inhibitor cellular automaton for the formation of patterns are presented. Our extended model allows studying the formatting of patterns and their temporal evolution also in the favourable and hostile environments. Particularly, its sensitivity to various initial conditions has been studied. Two different types of initial random configurations were taken into account: the uniform random distribution of differentiated cells (the R-system) and the non-uniform distribution in form of random Gaussian-clusters (the G-system). The most probable size of final stable population depends on the type of the initial configurations as well as the environmental conditions. The participation of a favourable environment is more clearly seen for the G-system. In addition, the G-system as being initially more disordered compared to the R-system usually evolves to a more spatially inhomogeneous final pattern. We show that each of the systems is subject to different dynamics. The results of the analysis shed also a light on some features in the evolving model such as the appearing of the oscillatory behaviour of the population size. Probably, this phenomenon has a connection with the impact of the favourable environment, which in a simple way was incorporated into our model. The more general conclusions could be obtained by consideration additional types of initial spatial distributions, possible various anisotropies in an environment as well as the asynchronous updating of a system. These suggestions can be interesting topics of a future study with regard to the current model.